\documentclass[aps,reprint,pra,floats,showpacs,amsmath,amssymb]{revtex4-2}
\usepackage{notoccite}
\usepackage{graphicx}  
\usepackage{dcolumn}   
\usepackage{placeins}
\usepackage{float}
\usepackage{longtable}
\usepackage{siunitx}
\usepackage{xcolor}
\usepackage{ulem}
\usepackage{soul}
\usepackage{url}
\usepackage{float}
\usepackage{datetime}
\usepackage{longtable}
\usepackage{ amsmath,accents}
\renewcommand{\v}[1]{\mathbf{#1}}

\def\ab{{\it ab-initio }}

\def\k{\v{e}_z}

\def\k{\v{k}}
\def\k0{\v{ k}_0}

\newcommand{\quotes}[1]{\lq\lq #1\rq\rq}

\def\deg{{}^{\circ}}

\def\odu{Physics Department, Old Dominion University, Norfolk,Virginia,USA}
\def\camchem{Yusuf Hamied Department of Chemistry, University of Cambridge, Lensfield Road, Cambridge CB2 1EW, UK}

\def\threeo2{\frac{3}{2}}

\def\colmang {\si{\angstrom}}
 
\renewcommand{\v}[1]{\mathbf{#1}}

\def\odu{Physics Department, Old Dominion University, Norfolk,
Virginia,23529, USA}

\begin{document}
\title{\color{blue}The density and pressure of helium nano-bubbles encapsulated in  silicon}
\def\dena{atoms$\;\colmang^{-3}$}
\def\sigmau{atoms$\;\colmang^{-1}\,$}
\def\colmang {\si{\angstrom}}
\author{N. C. Pyper}
\author{A.J.W. Thom}
\affiliation{\camchem}
\author{Colm T. Whelan}
\affiliation{\odu}
\email{cwhelan@odu.edu}

\begin{abstract}
The $1s^2 \rightarrow1s2p(^1P)$ excitation in confined and compressed helium atoms in either the bulk material or encapsulated in a bubble is shifted to energies higher than that in  the free atom. For  bulk helium, the energy shifts predicted from non-empirical electronic structure computations are in excellent agreement with the experimentally determined values. However, there are significant discrepancies both between the results of experiments on different bubbles and between these and the well established descriptions of the bulk.  A critique is presented of previous attempts to determine the densities in bubbles by measuring the intensities of the electrons inelastically scattered in STEM experiments. The reported densities are untrustworthy because it was assumed that the cross section for inelastic electron scattering was the same as that of a free atom whilst it is now known that this property is greatly enhanced for atoms confined at significant pressures. 
 It is shown how experimental measurements of bubbles can be combined with data on the bulk using a graphical method to determine whether the behavior of an encapsulated guest differs from that in the bulk material.  Experimental electron energy loss data from an earlier study of helium encapsulated in silicon is reanalyzed using this new method to show that the properties of the helium in these bubbles do not differ significantly from those in the bulk thereby enabling the densities in the bubbles to be determined. These enable the bubble pressures to be deduced from a well established experimentally derived equation of state. It is shown that the errors of up to 80\% in the incorrectly determined densities are greatly magnified in the predicted pressures which can be too large by factors of over seven.       
{\color{black}
This has major practical implications for the study of radiation damage of materials exposed to $\alpha$ particle bombardment}. 

\end{abstract}
\maketitle

 Helium bubbles are  formed   in materials subject to $\alpha$ particle radiation. Such bubbles are technologically significant because they  occur during ion implantation in Si \cite{,schierholz15a}, in nuclear waste disposal \cite{wang20} and  readily accrue in the metallic cladding of nuclear reactors \cite{zinkle13},
where they can  exert such pressures that the structural integrity of the host material is put at risk \cite{li19}. The problems caused by confined and compressed helium are fast becoming a major issue in nuclear fusion materials research \cite{zinkle13,hammond20}. Encapsulated  helium bubbles not only  cause embrittlement and fatigue \cite{cowgill20a} but also  damage  the  plasma-facing surfaces forming  coral like nano-structures which could erode into the plasma, degrading its quality, cooling and suppressing   the fusion reaction \cite{reinhart22a}.  

Scanning Transmission Electron Microscope(STEM) studies allow electronic excitations to be explored at high spatial resolution and
there have been numerous studies derived from electron energy-loss spectroscopy (EELS)  of helium nano bubbles  embedded in different  host materials \cite{jager82,walsh00,taverna08a,frechard09,david11,david14,alix15,sugar18,taylor22}. 
A knowledge of the density in a bubble is needed to deduce the pressure exerted on the boundary atomic matrix. 
\begin{table*}[t!]
\centering
\begin{tabular}{|r|c|c|c|c|c|c|c|c|c|l|}
\hline
\multicolumn{11}{|c|}
{Computed energy shifts, scattering cross-sections and $\Sigma$} \\ \hline
n            &0.04 &0.0419 \cite{arms01}  &0.0446 &0.045&0.046 \cite{arms05}& 0.0492\cite{arms05} &0.05&0.0518\cite{arms05}&0.0536 \cite{arms99}  &0.0562 \cite{arms01}  \\ \hline 
$\Delta E$&0.856 & 0.919 &1.006 &1.018 &1.052   &1.157 &1.182&1.242 &1.301&1.384 \\ \hline
$\sigma^{2p}_{fcce}(n)$ &0.581 &0.592 & 0.594&0.595 &0.606 &0.622 &0.625&0.630&0.634&0.649\\ \hline 
$\sigma^{2p}_{free}(n)$& 0.367  &0.366 &0.364&0.364&0.363&0.361&0.361&0.359&0.358&0.357\\\hline
$\Sigma$   &0.0232 & 0.0248& 0.0265 & 0.0268   & 0.0279 &0.0306&0.0313&0.0326&0.0340&0.0365  \\\hline            
\end{tabular}
\caption{Densities,$n$, are given in atoms $\colmang{}^{-3}$, cross sections, in $10^{-3}\colmang{}^2$, Energy shifts, $\Delta E$, in eV, $\Sigma$ in $10^{-3}$atoms $\colmang{}^{-1}$, citations are to the experimental measurements. All calculations, except free atom cross sections, in the fcce model of [21]
 \label{table1}}
\end{table*}

The results of such STEM experiments have been used in two different ways to try to determine the atomic densities in the helium bubbles. 
The first method for determining densities relies on the result that the lowest allowed electronic excitation is always shifted to energies greater than that of 21.218 eV. in the free atom, thereby raising the possibility that the density could be deduced by measuring this energy shift. In the second method for investigating bubble densities, the experimentally measured quantities are the bubble diameter and ratio of the numbers of inelastically and elastically scattered electrons \cite{walsh00}.  The density can then be deduced if the value of the scattering cross-section for one atom is known.
Unfortunately, the STEM energy shifts reported in the literature vary greatly in non-systematic ways both for bubbles in different hosts and between different experiments using the same host. 

Independently of the STEM studies, bulk helium confined and compressed under high pressures has been investigated in  very precise  inelastic X-ray scattering (IXS) \cite{arms99,arms01,arms05} and vacuum UV \cite{surko69} experiments. These revealed that  the $1s^2 $ to $1s2p(^1P)$ excitation energy increased systematically with density enhancement. This behavior is qualitatively similar to the non-systematic enhancements observed in the STEM experiments solely in that all the shifts are positive. For the bulk condensed phases,  the energy shifts and scattering cross-sections have been investigated theoretically \cite{pyper01,pyper17,pyper21,pyper22} by using an entirely non-empirical approach in which helium atom wavefunctions were computed as a function of atomic density. The density dependence of the predicted energy enhancements agreed very well with both those measured in the IXS and  and vacuum UV experiments in which the densities and pressures were determined completely independently, not invoking the energy shifts. However, these results stand in stark contrast to those of the  currently reported STEM studies for which the reported energy shifts are not readily understanable. 

The computed helium atom wavefunctions were also used \cite{pyper17,pyper22} to calculate the scattering cross-sections. It was found both that these increased monotonically with increasing density and that, for the densities considered \cite{arms01,arms05,arms99,pyper22},  the values were significantly greater than that for an isolated helium atom.  These results mean that the densities previously deduced by the second method of analyzing STEM measurements must be reexamined because it was assumed that the scattering cross-section remained essentially  unchanged from the free atom value.

 

The primary motivation for the microscope studies is the determination of the density and hence the pressure in encapsulated bubbles. However, in  neither of the methods used is the density explicitly measured, rather it is deduced from other measured quantities. To place the STEM results on a firm basis, a connection needs to be made to a situation   where both the properties of the helium condensed phase and the density are reliably known. The IXS measurements of \cite{arms99,arms01,arms05} provide energy shifts  at known pressures and densities and are thus ideal for this purpose.
The helium sample was contained in a high pressure system at a known temperature and pressure \cite{vent96}. The corresponding molar volumes were determined from existing $PVT$ measurements \cite{glassford66}. For each density, the measurements yielded a value for the $1s^2$ to $1s2p(^1P)$ excitation energy from which the energy shift $\Delta E$ can be derived by subtracting the free atom excitation energy of 21.218 eV. These experiments thus provide an accurate and entirely trustworthy set of data to compare with theoretical predictions.
This letter  has three objectives.
The first is to present a general method for analyzing experimental bubble measurements which will determine whether the behavior of the guest in a bubble is that same as that of the bulk guest. 
 The second is to reanalyze  the experiments of \cite{david11,david14,alix15,alix18} on  helium bubbles in silicon, in order to take account of the density dependence of the cross-sections. The third objective is to show
that the overestimations of the densities predicted using the free atom cross-section are greatly magnified in the predictions of the pressures. This has profound practical implication for the study of  $\alpha$ radiation  damage to materials.

In \cite{pyper21,pyper22} each energy shift was calculated as the sum of the prediction produced by an \ab self-consistent field (SCF) computation augmented by well defined non-empirical dispersion and short range correlation contributions.
The condensed helium phase was treated as  an fcc structured lattice using the extended model for the $1s2p(^1P)$ excited state which allows for the delocalization of the excitation \cite{pyper01,pyper21}. This approach will be denoted the fcce model. The $\Delta E$ values thus predicted \cite{pyper21} and presented in Table \ref{table1} are in almost perfect agreement with the experiments \cite{arms99,arms01,arms05}.  

 It remains  to forge a direct connection with the microscope experiments.
In the STEM experiments, the energy loss in the inelastic electron scattering as well as the bubble diameter $d$ and the ratio $I_p/I_z$
of the intensities of the scattered and incident beams are all  simultaneously measured.  The number density can then, in principle, be determined from the relation: \cite{walsh00, egerton96}  
 \begin{equation} \label{cs1}
n = \frac{ I_{p} }{ \sigma^{2p}(n) I_{z} d }.
\end{equation}
where $\sigma^{2p}(n)$  is the density dependent inelastic scattering cross-section for the $1s^2$ to $1s2p(^1P)$ excitation. 
The quantity $\Sigma$ defined by
\begin{equation}\label{sigma1}
\Sigma= \frac{ I_{p} }{I_{z} d }
\end{equation}
contains only explicitly measured quantities and can be directly compared with $n \sigma^{2p}(n)$ from theory.

In atomic units, the first Born cross section for the excitation of the target atom from the ground state to first excited state by an electron with an impact energy, $E_i$ is given by \cite{inokuti71,pyper22} 
\begin{eqnarray}\label{born1}
\sigma^{2p}(n)&=&\frac{4\pi}{E_i}\int_{q_{min}}^{q_{max}}\frac{|\epsilon_{2p}(q)|^2}{q^3}dq
\end{eqnarray}
where $q_{min}$ and $q_{max}$ are the minimum and maximum values of the momentum transfer as given in \cite{pyper22} and $\epsilon_{2p}(q)$ is the atomic form factor.
In the microscope experiments  considered here $E_i$ is 200keV for which (\ref{born1})
is certainly valid.
The evaluation of (\ref{born1}) requires prior knowledge of $\Delta E$ because this enters the calculation of the integration limits.
In earlier bubble studies \cite{walsh00,taverna08a,frechard09,david11,david14,alix15,alix18,sugar18}
it was assumed that a bubble density could be deduced by substituting into equation (\ref{cs1}), the experimentally measured value of $\Sigma$ with  $\sigma^{2p}(n)$ computed by evaluating $|\epsilon_{2p}(q)|$ using the free atom wavefunctions but with the integration limits in (\ref{born1}) derived from the experimentally measured values of  $\Delta E$.

However, it has been shown elsewhere \cite{pyper22} that, in the condensed phase, the $2p$ orbital is significantly compressed which causes both the form factor and scattering cross section to be significantly enhanced compared with the free atom values.
In Table \ref{table1} we compare the free atom cross sections with those calculated using the face centered cubic lattice-extended(fcce) description  of \cite{pyper21} for a 200 keV impact energy. Even allowing for the
change in the limits of integration in (\ref{born1}), the free atom cross sections are only marginally
different from the gas phase value (0.385 $\times10^{-3}\colmang^2$) over the range of densities considered
while the fcce values are greatly enhanced. 

{\color{black}
Reliable knowledge of the behavior of helium in the bulk can only be used
to draw conclusions from the results
of experiments on helium encapsulated in a bubble if it has
already been established that the behavior of
the material is the same in these two possibly different environments.
For the bulk, both $\Sigma$ and $\Delta E$ can be computed from the
density, $n$.
In the densities studied computationally, both $\Sigma$ and $\Delta E$
increase monotonically with $n$, establishing a one-to-one
correspondence between them for the bulk.
%
If for a bubble, both $\Sigma$ and $\Delta E$ are measured, then computations for the bulk material determine the unique density for which the bulk $\Delta E$ value is identical to that of the bubble. For this density, the $\Sigma$ value computed as $n\sigma^{2p}(n)$ using the bulk electronic wavefunctions can then be compared with the experimental bubble $\Sigma$ deduced as ${I_p}/(I_zd)$. If these two $\Sigma$ values agree, the bubble density would have been reliably determined with the bulk and bubble properties having the same density dependence. Any disagreement between these two $\Sigma$ values would suggest that the properties of the bubble were affected by the encapsulating matrix.
%

 In the event that experimental values of $\Sigma$ and $\Delta E$ were measured for a range of bubbles, the bubble and bulk data are most conveniently compared by plotting the experimental $\Delta E$ values against those for  $\Sigma$ derived via (2). If the bulk and bubbles behaviors were the same, the plot of the bulk computed $\Delta E$ against the bulk computed $\Sigma$ would coincide with the $\Delta E$ versus $\Sigma$ plot for the bubbles.  This method of analysis provides the best approach for comparing the bubbles and the bulk given that the bubble experiments are sufficiently challenging that the microscope data is subject to  significant experimental uncertainties.

The experimental study \cite{david11} of individual bubbles of helium in silicon can be analyzed using the graphical method just described.
{\color{black}
The  bubbles were systematically degassed
thus reducing the density within each individual bubble while keeping the diameter constant. After each reduction in density, the energy shift was measured.
The beam energy in the microscope was 200keV.
In line with the other STEM studies,
the density was then  estimated using (\ref{cs1}) taking the cross section $\sigma^{2p}(n)$ to be the free atom value, $\sigma^{2p}_{free}(n)$.
It is to be stressed that the density is not explicitly measured in the  STEM  experiments. However,  $\Sigma$,  composed of directly measured quantities, can be compared to the equivalent theoretical quantity $n\sigma^{2p}(n)$. Fortunately, the Poitiers group retained a record of 
the measured quantities and consequently could provide $\Sigma$ values \cite{david22}.  In \cite{alix18}  a systematic error  in the  experimental energy shifts was identified and  the measured  $\Delta E$  corrected by a factor $\delta E^{s}$ which was determined by requiring that the least dense bubble should have a zero energy shift. In the Supplementary Materials   it is shown that this approach leads to an overestimation of the  systematic error. A better estimate  is provided.
The individual bubbles are the same as in \cite{david11}. 
The experimental values of $\Delta  E$ plotted as a function of $\Sigma(\equiv{ I_{p} }/({I_{z} d }))$ are presented in Figure \ref{Sigmaexp}. The resulting curve is seen to be essentially the same as the plot of the computed $\Delta E$ as a function of the computed $\Sigma$}.
 It can, therefore, be concluded that the behavior of the helium bubbles in silicon at $97.15\deg K$ is essentially the same as that of the bulk material. This allows the densities and pressures in these bubbles to be elucidated. 

 \begin{figure}[t!]
    \includegraphics[width=\linewidth,height=7.2cm]{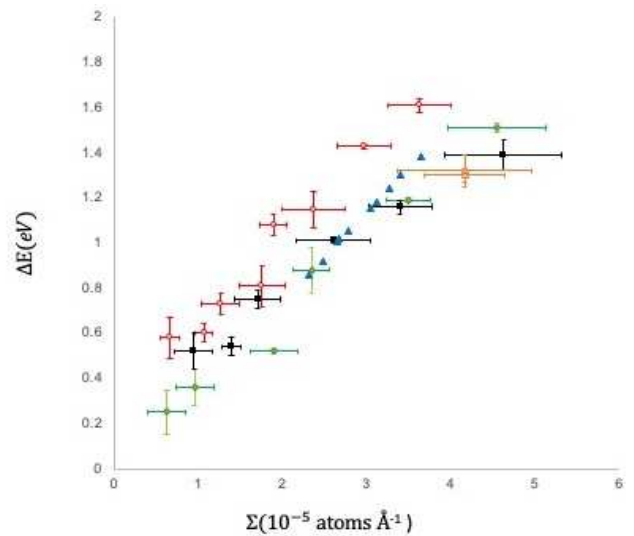}
     \caption{(Color on line) energy shift versus $\Sigma$ for the transition $1s^2\rightarrow 1s2p ({}^1P)$ . Shown are the theoretical results in the fcc- extended model of
    \cite{pyper22},
       blue filled triangles.
 Experiment [30] : bubble 1, red open circle; bubble 2,
solid green disk; bubble 3, black solid  rectangle; bubble 4, orange open triangle. The systematic error in the experimental determination of the bubble $\Delta E$ is taken as  our  best estimate of $\delta E^s= 0.16 eV$, as discussed in the supplementary material. \label{Sigmaexp}}
\end{figure}
\FloatBarrier

%
 {\color{black} 
 
 The pressures exerted by the helium atoms, one of the properties of primary interest, must be deduced  from the atomic densities by using an independently derived equation of state, preferably one obtained directly from experiment. 
 For our purposes the equation of state presented as equation (7) of Mills {\it et al}  \cite{mills80}  is 
 a suitable choice. It was parameterized from experimental measurements of all variables, the pressure, density and temperature.  
The  Mills equation  was experimentally validated for the temperature range from $75\deg K$ to $300\deg K$ and pressures from 200MPa to 2000MPa. 
  It is instructive to compare the exact densities and pressures  with those  which could be estimated  for a measured $\Sigma$.
  The $\Sigma$ values  given in Table \ref{table1} correspond to  exactly known densities.
Consider the maximum and minimun values of $\Sigma$ as given in the table, for the smallest  $\Sigma=0.0232\times10^{-3}$ \sigmau,  the exact density is 0.04 \dena  but  the {free atom} estimate is  0.06328 \dena  an overestimation of $58\%$.
Using the Mills equation we  have calculated the exact and free atom pressures. The free atom estimate is 648 MPa while the exact value is only 170 MPa.
The overestimations are ever more severe for the largest  $\Sigma=0.0365\times 10^{-3}$ \sigmau , the free atom approximation yields a density of 0.10233 \dena  which is 80\% larger than the exact value of  0.0562 \dena . Further, the  free atom pressure of 3328 MPa, is 7.5 times bigger than the exact value of 443 MPa.
The left panel in Figure 2 presents  the curve(solid green) relating the correct values of $\Sigma$ to the exactly corresponding densities as in Table \ref{table1}, while, in the right panel, the resulting pressures are shown (solid  green line) as a function of $\Sigma$. The  dashed curves in red in the two panels of Figure 2 show the significant overestimations of the densities and pressures that would be predicted if the free atom cross sections were used in equation (\ref{cs1}) in deriving the densities from the $\Sigma$ values. The pressures were computed for the 97.15$\deg K$ temperature of the bubble experiments \cite{david11,david22}. 
\begin{figure}[t!]
\centering
\includegraphics[width=\linewidth,height=6.0cm]{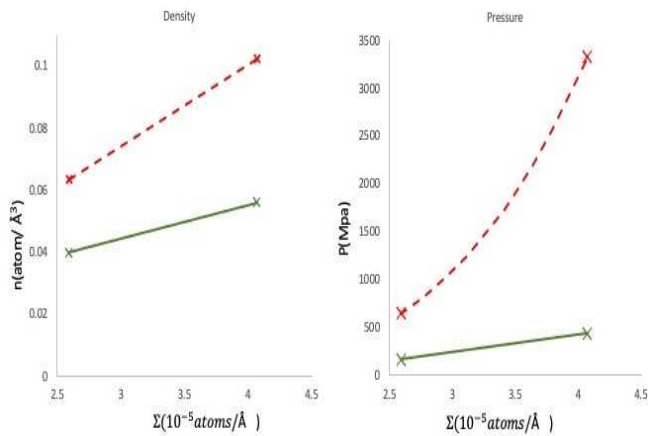}
\caption{(Color on line) Comparison of the exact  and free atom values for density and pressure. Left panel: density. Right panel: pressure. Green solid   line confined  values, red dashed  line free atom. The crossed points are for the two values, $\Sigma=0.0232\times10^{-3}$\sigmau
and $\Sigma=0.0365 \times10^{-5}$ \sigmau  highlighted in the text\label{denp}}
\end{figure}

We have concluded from the analysis of the data presented in Figure 1 that the behavior of helium bubbles in silicon is essentially the same as that in the bulk. Hence the 1:1 correspondence between the density, $\Sigma$ computed as $n \sigma^{2p}(n)$ and the pressure demonstrated for the bulk is also valid if $n \sigma^{2p}(n)$ is replaced by the value determined experimentally via equation (2). This allows the bubble pressures to be deduced from the experimental $\Sigma$ values. 
The $\Sigma$ values in Table \ref{table1} overlap with a large portion of the range of values measured for the encapsulated bubbles \cite{david11,david22}. 
Two of these $\Sigma$ 
values, namely $0.0235 \times 10^{-3}$ \sigmau and $ 0.0363 \times10^{-3}$ \sigmau, measured in the bubble experiments are virtually identical to the two values of $ 0.0232 \times10^{-3}$  \sigmau and $0.0365 \times10^{-3}$ \sigmau considered above. This shows that the pressures in these two bubbles will hardly deviate from those of 170 MPa and 443 MPa 
deduced above from the two very close $\Sigma$ values in Table \ref{table1}. Consequently, the pressures in the bubbles having $\Sigma$ greater than $0.0232 \times 10^{-3}$\sigmau will fall in the MPa range rather than that of GPa that would be deduced using the free atom cross sections in equation (\ref{cs1}).

In this letter we have focused on experiments  on helium bubbles encapsulated in  silicon  \cite{david11,david22}. 
The results presented here suggest that the relation between energy shift and density given in \cite{pyper22} is not sensitive to the  silicon host material, being essentially the same in the bubble and the bulk.  Silicon is  a semi-conductor, further work  is needed to see if this is also true for metallic hosts.

In virtually all 
STEM studies \cite{walsh00,david11,david14,alix15,taverna08a,frechard09,ono19,evin21} the density has been estimated using the free atom $\sigma^{2p}_{free}(n)$. This method of attempting to deduce the density is not correct and will not yield trustworthy values because the free atom cross section  is much smaller than that of an atom in the condensed phase. We have presented a general method for determining whether the behavior of a helium atom encapsulated  in a bubble is the same as that in the bulk.
 It has been  demonstrated that the free atom  approach to determining the density grossly overestimates both it and the pressure. This has very significant implications for the study of materials exposed to  $\alpha$  particle bombardment.}

We are grateful to Dr Marie-Laure David for supplying us with her experimental data and for very many helpful discussions
\newpage
{\bf Supplementary Material:Determination of the systematic experimental error}
%
\FloatBarrier
\begin{figure}[H]]
\includegraphics[width=\linewidth, height=5.6cm]
{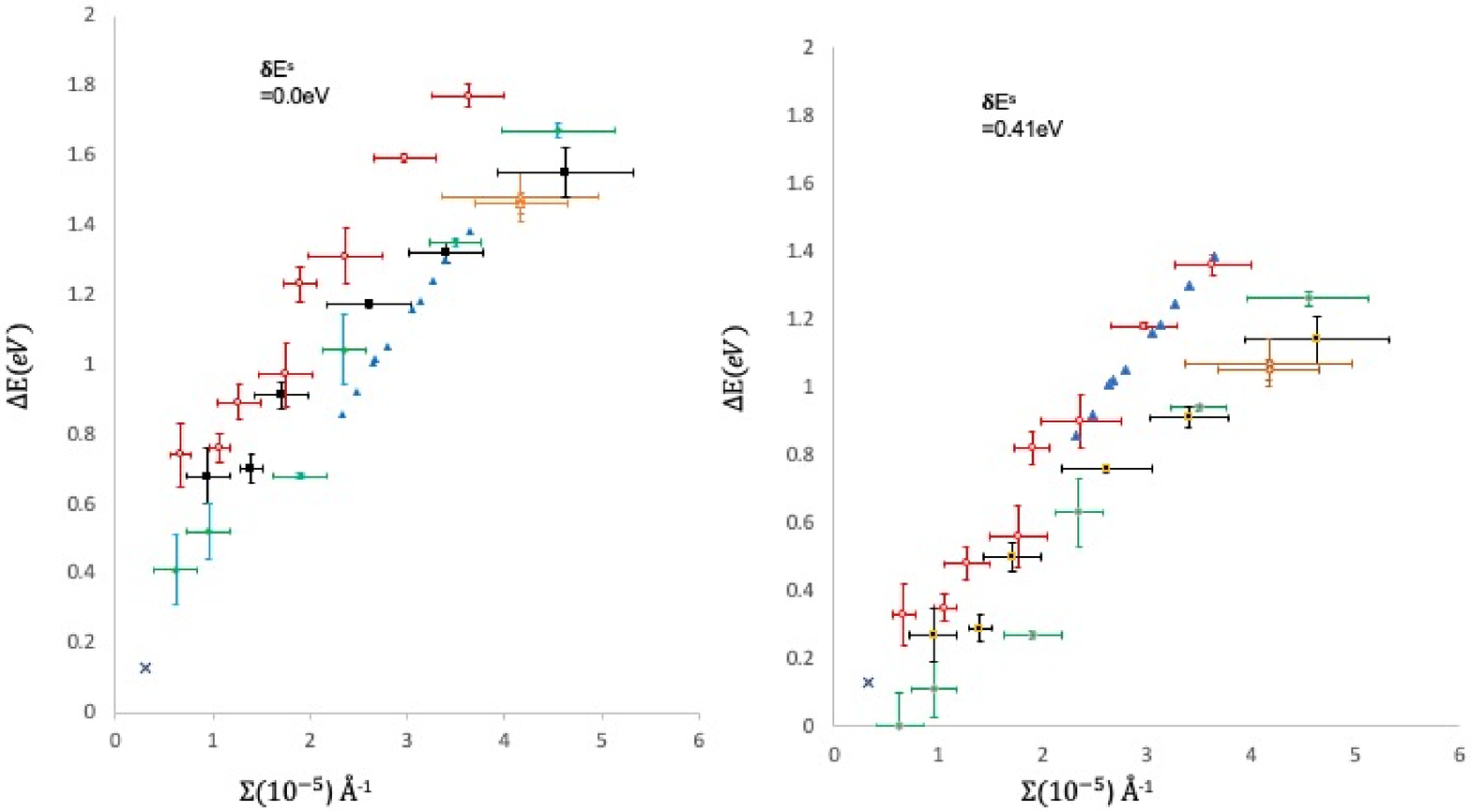}   
   \caption{ Energy shift($eV$) versus $\Sigma$ for the transition $1s^2\rightarrow 1s2p ({}^1P)$. Left  panel: $\delta E^{s}=0.0eV$
    Right panel: $\delta E^{s}=0.41eV$. Shown are the theoretical results in the fcc-extended model of \cite{pyper22}, blue filled triangles; theoretical reference point, cross.
    Experiment: bubble 1, red open circle; bubble 2, green open diamond; bubble 3, black solid rectangle; bubble 4, orange open triangle. \label{SMone}}
    \end{figure}
\FloatBarrier

{\color{black}
The quantities measured in the STEM/EELS experiments are $\Sigma \; [=I_p/(I_zd)]$ and values for the $1s^2 \rightarrow1s2p(^1P)$ excitation energy, conventionally known as the K edge. The energy shift $\Delta E$ can then, in principle, be derived by subtracting the free atom excitation energy of 21.218eV. The STEM data can  then analyzed from graphs of $\Delta E$ plotted against $\Sigma$. 
 However,  the observed K-edge is sensitive  to  the resolution  and  other parameters specific to a given  microscope \cite{alix18,pyper17}.
The experimental uncertainties in the current STEM measurements \cite{alix18,david11,david22} indicate that the true energy shifts, $\Delta E$, might deviate from the experimentally determined values, $\Delta E^{obs}$, by a systematic error, $\delta E^s$, so that }
\begin{eqnarray}\label{A1}
{\Delta E}^{obs}&=&\Delta E +\delta E^{s}.
\end{eqnarray}

Here,  we  discuss the determination of $\delta E^{s}$ for the specific experiments on silicon \cite{david11,david22} and suggest an appropriate value for $\delta E^{s}$.

As first step, we can set upper and lower bounds on this error.
 In the left panel of Figure  \ref{SMone}, we show the results with $\delta E^s$ assumed to be zero. This is, of course, the lower bound on $\delta E^s$. For bubble 2, the smallest value of
$\Delta E^{obs}$ corresponds to a density sufficiently low that it could be reasonably assumed that the true energy shift is tending to zero. If it is assumed that this is exactly zero,  we arrive at a value of
$\delta E^s=0.41$ eV in agreement with \quotes{corrected} energy shifts given in \cite{alix18}.  The graph presented in the right hand panel of Figure  \ref{SMone} is constructed by taking $\delta E^{s}$ to be 0.41 eV. As $n \rightarrow 0 $, not only does $\Delta E \rightarrow 0$, but $\Sigma $ $ [= n\sigma^{2p}(n)] $ also tends to 0 because $\sigma^{2p}(n)$ tends to the free atom value. However, from  the graph we find  a finite value for $\Sigma$ at  $\Delta E = 0$. It follows that 0.41 eV. is too large a value for the systematic error correction $\delta E^s.$ In summary
\begin{eqnarray}\label{A2}
0.0\le \delta E^{s}< 0.41eV
\end{eqnarray}

Further progress requires a theoretical investigation of a very low density bubble for which both $\Delta E$ and $\Sigma$ are tending to zero. The experimentally determined pair distribution function for the liquid at density 0.022 \dena, experiencing its ambient vapour pressure, showed that the environment of each atom was close to that in a bcc structured lattice. Furthermore, this distribution function, depicted in figure 6 of \cite{lucas83}, showed  a very clear first coordination shell of eight atoms but with the second shell much less well defined. These results suggest that, for the liquid at the even lower density of 0.0083 \dena the environment of any one atom is best modelled by considering its interactions with just a single coordination shell of eight atoms. For this system, we computed an energy shift $\Delta E$ of 0.13eV. by using the previously described \cite{pyper21} non-empirical electronic structure calculations.This density is sufficiently small that the cross section has gone over to the free atom value of $0.385\times 10^{-3}\colmang^2$ and thus $\Sigma\approx0.32\times 10^{-5}$ atoms  $\colmang^{-1}$.
This yields a  small $\Sigma$, small $\Delta E$ reference point shown as a cross in Figure \ref{SMone}.

    {\color{black}

In Figure \ref{SMtwo}, the  measurements for the different bubbles are plotted for $ \delta E^{s} $ taken to 0.16 eV and 0.3 eV. Bubble 2 has the lowest density. 
For the larger systematic error  the reference point lies close to the  edge of the horizontal error  bar for  the smallest  energy shift of  bubble 2. However, for $\delta E^{s}= 0.16 eV$ the reference point and the smallest two experimental points all lie on a single line which passes through the origin suggesting a smooth transition to the correct limiting value as $n\rightarrow 0$. Consequently, we feel that $\delta E^{s}=0.16eV$ is the  best estimate of  the systematic error. 

\begin{figure}[t!]
\includegraphics[width=\linewidth, height=6.0cm]{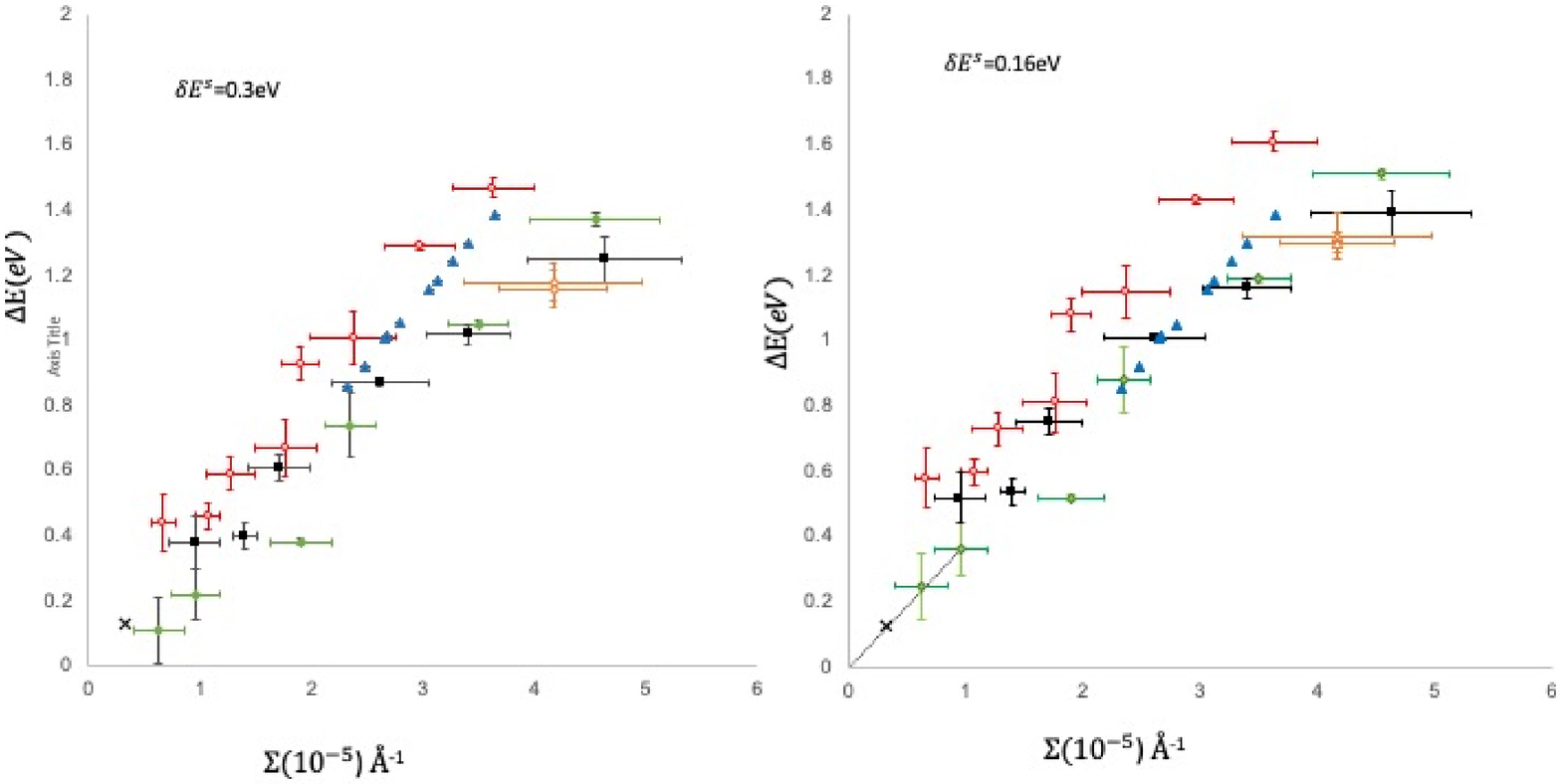}  
\caption{Energy shift versus $\Sigma$ for the transition $1s^2\rightarrow 1s2p( {}^1P)$. Left  panel: $\delta E^{s}=0.3eV$.
    Right panel:  $\delta E^{s}=0.16eV$.  Symbols as in Figure \ref{SMone}
\label{SMtwo}}
   \end{figure}
    \FloatBarrier
    \newpage
There are sufficient inconsistencies between the individual bubble measurements to suggest that the uncertainties in the measured $\Sigma$'s are much larger than the error bars shown.
It follows that
the fcee calculations are not inconsistent with the results for the full range of possible systematic errors.
   

%

\bibliographystyle{apsrev4-2}
\bibliography{confined_mat}{}

\end{document}